\documentclass[showpacs,amsmath,amssymb,twocolumn,prl,superscriptaddress]{revtex4}

\usepackage[dvips]{graphicx} 
\usepackage{amsfonts}
\usepackage{amscd}
\usepackage{amsmath}    
\usepackage{enumerate}

\begin{document}

\title{Quantum key distribution secure against the efficiency loophole}

\author{Xiongfeng Ma}
\email{xfma@iqc.ca}
\author{Tobias Moroder}
\author{Norbert L\"utkenhaus}
\affiliation{%
Institute for Quantum Computing and Department of Physics and Astronomy, \\
University of Waterloo, 200 University Ave W., Waterloo, ON, Canada N2L 3G1 \\
}%

\begin{abstract}
An efficiency-loophole-free quantum key distribution (QKD) scheme is proposed, which involves no hardware change but a modification in the data post-processing step. The scheme applies to a generic class of detection systems which allow correlations between the detection efficiency and the outcomes. We study the system transmittance and the detection error rate that allow  implementing this scheme. We also consider the case that a quantum memory is used to boost the performance and investigate the criteria of the readout probability. The simulation result shows that the lowest tolerable system transmittance (readout probability) is 50\%, while the necessary efficiency for an efficiency-loophole-free Bell's inequality test is 82.8\%.
\end{abstract}

\maketitle

Bell's inequalities \cite{Bell_Ineq_64} play a critical role in testing against alternative theories of quantum mechanics, like the local hidden variable theory. Following the early Bell test experiments \cite{Freedman.Bell.72,Aspect.Bell.81}, there are many experiments performed in the last decade \cite{Tittel.Bell.98,Weihs.Bell.98,Pan.BellGHZ.00,Rowe.Bell.01} that have favored quantum mechanics. However, the locality loophole \cite{Bohm.Local.57} and the efficiency loophole (or fair sampling problem) \cite{Pearle_Bell_70} block these experiments to draw a decisively conclusion. Note that no Bell test experiment has been done so far to close all these loopholes simultaneously.

The efficiency loophole exists when the data post-processing only focuses on a certain part, but not all, of the prepared states. For instance, in many Bell test experiments (especially for optical implementations), data used to test the inequality are conditional on coincidence events, where detections occur in both wings of the experiment. The underlying assumption for this post-selecting is that the sample of detected states is representative of all the states prepared. Unfortunately, this fair sampling assumption can not be tested (proved) from the experiment. This is normally called detection efficiency loophole or fair sampling problem.

Since the concept of entanglement is closely related to Bell's inequalities \footnote{Although entanglement does not promise violation of the Bell's inequality.} and entanglement is the precondition of quantum key distribution (QKD) security \cite{CLL_Precondition_04}, a natural question is ``Does this efficiency loophole affect the security of QKD?" The answer is \emph{yes}.

In theory, the security of QKD has been proven in literature \cite{Mayers_01,LoChauQKD_99,ShorPreskill_00}. When it comes to real-life implementations, various device imperfections should be taken into consideration. A lot of efforts have been made to achieve the security of QKD with realistic devices \cite{KoashiPreskill_03,GLLP_04}. Nevertheless, not all the imperfections have been fully examined yet. Hence, these security analyses cannot be applied to some realistic cases. For example, the efficiencies of two detectors (one detects 0 and the other detects 1) in a QKD system may not be the same, which opens up a loophole to attacks such as the efficiency mismatch attack \cite{MAS_Eff_06} and the time-shift attack \cite{QFLM_TimeShift_07}. Note that the time-shift attack has been successfully demonstrated in lab recently \cite{Timeshift_Exp_08}. A simple counter measure on the single photon level to these two attacks exists by randomly switching two detectors \cite{Nielsen_Fourstate_01,Mismatch_security_08}.
However, this counter measure does not close the efficiency loophole and further attacks can be launched \cite{QFZMTCL_attack_07}. All in all, the system transmission efficiency opens a loophole to QKD attacks. This loophole stems from the aforementioned fair sampling problem in Bell's inequality test.

In this Letter, we examine a typical detection system and explain why the efficiency loophole exists in QKD. We then propose an efficiency-loophole-free QKD scheme. The proposed scheme differs from the existing QKD schemes only in the classical data post-processing procedure, mainly the privacy amplification part. No hardware change is required for the new scheme. Finally, we apply it to various photon sources with and without quantum memory to show the criteria for QKD devices to implement this scheme.

There are two types of QKD schemes. In the prepare-and-measure scheme, Alice, the sender, prepares a state and sends it to Bob, the receiver. In the entanglement based scheme, an eavesdropper, Eve, prepares a bipartite state for Alice and Bob. In the security proof, we assume that Eve controls the channel. Thus, in either cases, the state to be detected is prepared by Eve. In general, the state coming to a detection system can be in an arbitrary-dimensional Hilbert space. The following discussion will focus on the prepare-and-measure case, where we only need to consider the detection system on Bob's side. For the entanglement based scheme, Alice's detection system is the same as Bob's.


In the following discussion, we will focus on a generic case that the detection system is composed of two detectors (for bit 0 and 1). Then, in a real experiment, there are four possible outputs: neither of the detector clicks (no click), one detector clicks (0), the other detector clicks (1) and both detectors click (double click). A traditional data pre-processing works as follows: discard all no clicks, keep single clicks and randomly assign 0 or 1 to double clicks.

In theory, a so-called squashing model is widely used in security proofs of QKD \cite{KoashiPreskill_03,GLLP_04}. Note that security proofs of QKD with a threshold detector model \footnote{A threshold detector can only tell whether the input signal is vacuum or not.}, are presented \cite{Koashi_NewModel_06,TT_Thres_08,BML_Squash_08} recently.

\begin{figure}[hbt]
\centering \resizebox{6cm}{!}{\includegraphics{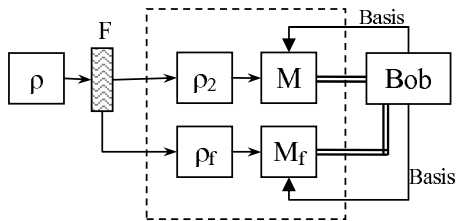}}
\caption{A schematic diagram for a detection system. The operator ($F$) is the key component of the squashing model, which projects the incoming state $\rho$ into a qubit state $\rho_2$ and a flag state $\rho_f$. The measurement outcome of the flag state $\rho_f$ gives Bob information about no/single/double click.} \label{Fig:Squash}
\end{figure}

\textbf{Squashing model:} A detection system first performs an operation $F$ mapping the incoming state $\rho$ (in an arbitrary-dimensional Hilbert space) into a two-dimensional pace (qubit) state $\rho_2$ with an ancillary state $\rho_f$ (in an arbitrary-dimensional space). Then two measurements are performed on $\rho_2$ with outcome 0/1 and on $\rho_f$ with outcome of a flag showing \emph{no/single/double} click. 
Note that both measurements of $M$ and $M_f$ may depend on the basis choice. The schematic diagram is shown in Fig.~\ref{Fig:Squash}.

Typical security proofs of QKD rely on the assumption that the measurement outcome on $\rho_f$ is independent of the squashed qubit $\rho_2$. With this assumption, Alice and Bob can safely discard all no clicks and focus on other detections in the post-processing. In reality, this assumption might not be valid. One example is the time-shift attack \cite{QFLM_TimeShift_07}. Similar to Bell's inequality tests, this is where the efficiency loophole (or fair sampling problem) comes into the scenario.

One way to close the efficiency loophole in QKD is by making the overall transmittance to be 100\%. This is not feasible with current technology. From the discussion above, we already know that the fair sampling problem exists in QKD because of discarding no clicks in the data pre-processing. In order to have an efficiency-loophole-free QKD scheme, we only need to modify the data pre-processing: keep single clicks and randomly assign 0 or 1 for no clicks and double clicks. Now that the data from all states are used for the post-processing, the fair sampling problem is avoided. 


With this modified pre-processing, Bob always detects 0 or 1 for each pulse sent by Alice. It is equivalent to say that the system transmittance is 100\%. On the other hand, Bob has extra information from the measurement on the flag state $\rho_f$. Shortly, we will see that this information is useful for key generation. Now, the definitions of gain and quantum bit error rate (QBER) are different from the traditional ones, where the data post-processing only focuses on the single and double clicks. Define $Q_s$ to be the rate for Bob to get a single click. Define $E_s$ to be the error rate given that Bob gets a single click. 
Note that both quantities, $Q_s$ and $E_s$, can be experimentally measured or tested.
The QBER with the new pre-processing is given by
\begin{equation} \label{Eff:QE}
\begin{aligned}
\delta &= E_sQ_s+e_0(1-Q_s),
\end{aligned}
\end{equation}
where the second term is due to the random bit assignment for the double clicks and no clicks and $e_0=1/2$. There are two types of errors: system errors, with a rate of $E_s$, and the ones caused by random bit assigning, with a rate of $e_0=1/2$.

\textbf{Basis independent source:} There are two types of basis independent sources \cite{KoashiPreskill_03}: perfect single photon source and entangled photon source \cite{EntanglementPDC_07}. For a QKD system using a basis independent source, the phase error probability is the same as the bit error probability,
\begin{equation} \label{Eff:PhaseInd}
\delta_p=\delta_b=\delta,
\end{equation}
where the QBER $\delta$ is given by Eq.~\eqref{Eff:QE}. Here we consider the infinite key length limit. Then the key rate is given by \cite{KoashiPreskill_03}
\begin{equation} \label{Eff:KeyInd0}
R \geq 1-H_2(\delta)-H_2(\delta),
\end{equation}
where $H_2(x)=-x\log_2(x)-(1-x)\log_2(1-x)$ is the binary entropy function.

As mentioned above, in the efficiency-loophole-free QKD scheme, Bob randomly assigns 0 or 1 for no clicks and double clicks. One observation is that Bob knows the location of the random bits. In the squash model we described in Fig.~\ref{Fig:Squash}, this location information comes from the measurement on the flag state $\rho_f$. Hence, he can group the key bits into two strings: the one coming from single clicks ($s_s$) and the other from random assignments ($s_r$). Alice and Bob together can experimentally test the error rate in $s_s$, which is $E_s$. Suppose the phase error rates in $s_s$ and $s_r$ are $E_{ps}$ and $E_{pr}$, respectively, then the overall phase error probability $\delta_p$ can be expressed by (in the long key limit), $\delta_p = Q_sE_{ps}+(1-Q_s)E_{pr}$.
Due to the fact that $0\le Q_s\le1$ and $E_{pr}\ge0$, we can have the upper bound of the phase error rate in $s_s$,
\begin{equation} \label{Eff:PhaseSingle}
\begin{aligned}
E_{ps} \le \frac{\delta_p}{Q_s} = \frac{\delta}{Q_s}.
\end{aligned}
\end{equation}
Thus the key rate from the single click part is given by
\begin{equation} \label{Eff:KeyInd1}
\begin{aligned}
R &\geq  Q_s[1-H_2(E_s)-H_2(\frac{\delta}{Q_s})], \\
\end{aligned}
\end{equation}
where $\delta$ is given by Eq.~\eqref{Eff:QE}. We know that no positive key can be extracted from the random bit string. So Eq.~\eqref{Eff:KeyInd1} gives a lower bound of the overall key rate. Thus, in the post-processing, Alice and Bob discard all no clicks and double clicks, and perform error correction and privacy amplification on the bits from single clicks according to Eq.~\eqref{Eff:KeyInd1}. From here we can see that the new scheme differs from the traditional one in two aspects.
\begin{enumerate}
\item
Double clicks are discarded in the new scheme, while they are randomly assigned to be 0/1 in the traditional scheme.

\item
More privacy amplification is performed in the new scheme according to Eq.~\eqref{Eff:KeyInd1}.
\end{enumerate}

To illustrate the intuition behind Eq.~\eqref{Eff:KeyInd1}, let us first consider two attacks with a simple case where $E_s=0$. When $E_s=0$, we can directly conclude from Eq.~\eqref{Eff:KeyInd1} that $Q_s>0.5$ is the lowest tolerable single click rate.

\textbf{The extreme case of time-shift attack} \cite{QFLM_TimeShift_07}: Eve can control detectors to be active (100\% efficiency) or inactive (0\% efficiency). A simple attack is that Eve randomly chooses one of the two detectors to be active and the other inactive. In this case, Eve knows all the information about the key and introduces 50\% loss, which follows $Q_s\le0.5$. No positive key can be obtained from Eq.~\eqref{Eff:KeyInd1} in this case.

\textbf{The strong pulse attack} \cite{Lutkenhaus_99DoubleClick}: Eve randomly chooses a basis to measure the state sent by Alice and replace it with a strong pulse according to the measurement result. For example, Eve chooses $X$ basis measurement and obtains an outcome of $+$. She resends many copies of $+$ state. In this case, Bob gets either a $+$ outcome (if he chooses $X$ basis measurement) or a double click (with a high probability if he chooses $Z$ basis measurement) with equal probabilities. Again, in this case, we get $Q_s\le0.5$, which leads no positive key from Eq.~\eqref{Eff:KeyInd1}.



Since the lowest tolerable transmittance (roughly $Q_s$) is around $50\%$ and we know that the dark count rate is low, typically below $10^{-4}$, in the following simulations, we neglect the dark counts.

\begin{figure}[hbt]
\centering \resizebox{8cm}{!}{\includegraphics{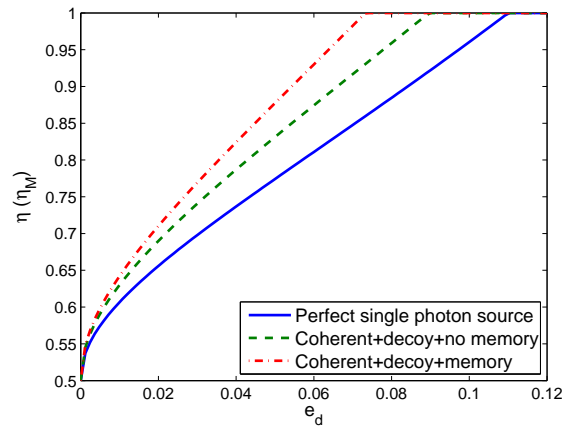}} \caption{Plot of the lower bound of the tolerable $e_d$ and $\eta$ ($\eta_M$) for the efficiency-loophole-free QKD scheme. For the two cases with coherent states, we pick up $\mu=0.5$ for the simulation. For the case of Coherent+decoy+memory, we assume the channel transmission is $\eta_c=1\%$.} \label{Fig:Eff:Lower}
\end{figure}

Now, let us study a simple case that Alice uses a single photon source. From Eq.~\eqref{Eff:QE}, the single click rate and its error rate are given by
\begin{equation} \label{Eff:Singlepara}
\begin{aligned}
Q_s &= \eta, \\
E_s &= e_d, \\
\delta &= e_d\eta+e_0(1-\eta), \\
\end{aligned}
\end{equation}
where $\eta$ is the overall transmittance, taking into account the channel loss and the detection efficiency, and $e_d$ is the intrinsic detection error probability. Now we can see that there are two crucial QKD system parameters, $e_d$ and $\eta$, determining the key rate given by Eq.~\eqref{Eff:KeyInd1}. The lower bound for the tolerable $e_d$ and $\eta$ is shown in Fig.~\ref{Fig:Eff:Lower}.

A parametric down-conversion (PDC) source is widely used as an entangled photon source, which is another type of basis independent source. Following the model presented in Ref.~\cite{EntanglementPDC_07}, one can calculate the single click rate, its error probability and then the key rate by Eq.~\eqref{Eff:KeyInd1}. We have numerically checked all possible choice of the source intensities $\mu$, but no positive key rate $R$ by Eq.~\eqref{Eff:KeyInd1} can be achieved. 

\textbf{Coherent state source:} Now, let us consider another case that a weak coherent state photon source is used.
Here we apply the GLLP security analysis \cite{GLLP_04} to the decoy state QKD scheme \cite{Hwang_03,Decoy_05}.
Similar to the derivation of Eq.~\eqref{Eff:KeyInd1}, the key rate of the efficiency-loophole-free QKD scheme with a coherent state source is given by
\begin{equation} \label{Eff:KeyCoh1}
\begin{aligned}
R &\geq
-Q_sH_2(E_s)+P_1Y_1[1-H_2(\frac{\delta_1}{Y_1})], \\
\end{aligned}
\end{equation}
where $Y_1$ is the probability that Bob gets a single click given Alice sends out a single photon state, $P_1=\mu e^{-\mu}$ is the probability Alice sends out a single photon state, and 
$\delta_1$ is the error rate of the single photon states including random bit assignment. The single photon state yield $Y_1$ and its error rate $e_1$ can be estimated by decoy states. Assume that the overall transmittance is $\eta$, then the parameters $Y_1$ and $\delta_1$ are given by
\begin{equation} \label{Eff:Cohreal}
\begin{aligned}
Y_1 &= \eta, \\
\delta_1 
&= e_dY_1+e_0(1-Y_1), \\
\end{aligned}
\end{equation}
Here in the simulation, we neglect dark counts and double clicks, so that $Q_s=Q_\mu=1-e^{-\eta\mu}$ and $E_s=E_\mu=e_d$. We can see that when $\mu$ approaches to 0, Eq.~\eqref{Eff:KeyCoh1} will converge to Eq.~\eqref{Eff:KeyInd1} except for an overall ratio. To illustrate the criteria for case of decoy state QKD with coherent state source, we pick up a typical value of $\mu$ used in real experiments, $\mu=0.5$, for the simulation. 
Now, we can find the relation between the tolerable $e_d$ and $\eta$, as shown in Fig.~\ref{Fig:Eff:Lower}. 


From Fig.~\ref{Fig:Eff:Lower}, we know that the transmittance, $\eta$, is required to be higher than 50\% for a QKD system even when a perfect single photon source is used. With the current (optical) technology, it is not feasible yet. One way to solve this problem would be to perform a quantum non-demolition (QND) measurement, distinguishing vacuum and non-vacuum states, right before Bob's box. This QND measurement gives Bob a trigger when the incoming signal is a vacuum or not. Since Bob chooses the basis (randomly) after the QND measurement, the security only relies on the state conditional on the trigger. Note that this QND measurement can also help close the efficiency loophole in the Bell's inequality test.


One candidate for this QND measurement is using a quantum memory. The requirements for a quantum memory to be used in the efficiency-loophole-free QKD scheme are listed as follows.
\begin{enumerate}
\item
The quantum memory can tell whether a photon is stored or not. When a photon is stored, a trigger will be emitted. 

\item
Given that the quantum memory emits a trigger, the readout probability is $\eta_M$, which takes into account the detection efficiency and the mis-trigger probability. The quantum memory is not required to re-emits photons for readout. A measurement can be performed on the quantum memory directly.

\item
The user can choose measurement basis after the quantum memory sends out a trigger. Then the security only relies on the quantum state stored in the quantum memory. Thus, only the readout efficiency (normally less than 100\%) will cause the random sampling problem.
\end{enumerate}

We remark that the quantum relay \cite{DLCZ_01} can be used for the QND measurement as well.




In the case that a basis-independent source (e.g., an entangled PDC source) is used, we can use Eq.~\eqref{Eff:Singlepara} in Eq.~\eqref{Eff:KeyInd1}. In this case, $\eta$ is replaced by the readout probability of the quantum memory, $\eta_M$. The tolerable $\eta_M$ and $e_d$ will be the same as the ``Perfect single photon source" curve shown in Fig.~\ref{Fig:Eff:Lower}.


%
%
%
%

For the case that a weak coherent state photon source is used, we have
\begin{equation} \label{Eff:CohQM}
\begin{aligned}
P_1 
&= \frac{\eta_c\mu e^{-\mu}}{1-e^{-\eta_c\mu}}, \\
Q_s &= Y_1 = \eta_M, \\
E_s &= e_d, \\
\delta_1 &= e_d\eta_M+e_0(1-\eta_M), \\
\end{aligned}
\end{equation}
where $\eta_c$ is the channel transmittance, or the probability of the quantum memory to send out a trigger signal given that Alice sends out a photon. One can use Eq.~\eqref{Eff:CohQM} in Eq.~\eqref{Eff:KeyCoh1} to get the final secure key rate. Here the definitions of parameters, $\delta_1$, $P_1$ and $Y_1$, are different from the case without a quantum memory. In this case, these probabilities are conditional on the trigger sent by the quantum memory. Since they have close physical meanings, we will use the same set of notations for simplicity. In a real experiment, $R$, $\delta_1$, $P_1$ and $Y_1$ can be estimated by decoy states.
In the simulation, we use $\eta_c=1\%$, corresponding to a 100 km fiber link loss, and $\mu=0.5$. The result is shown in Fig.~\ref{Fig:Eff:Lower}.

%


%

In Fig.~\ref{Fig:Eff:Lower}, there is no positive key in the regime $E>11.0\%$ for the single photon source case, which is consistent with the result in the literature \cite{ShorPreskill_00}. Note that the y-axes are different for the Coherent+decoy+no memory and Coherent+decoy+memory curves. 

Conclusion remarks:
\begin{enumerate}
\item
One can also search for better post-processing schemes, such as those based on two-way classical communications in the post-processing \cite{TwoWay_03, TwoWay_06}. An interesting question is whether the transmittance $\eta>50\%$ is a hard limit for efficiency-loophole-free QKD schemes.



\item
Proving the equivalence between the squash model presented in Fig.~\ref{Fig:Squash} and the threshold detector model, like Ref.~\cite{BML_Squash_08}, is an interesting prospective topic.

\item
The idea of discarding random assignment bits and doing more privacy amplification can be applied to the normal QKD scheme where random bits are assigned for double clicks.

\item
Although the scheme is proposed for QKD usage, the post-processing idea may also be useful for efficiency-loophole-free Bell's inequality test, where the necessary efficiency is 82.8\% \cite{Mermin_Bell_86}. This efficiency criterion for Bell's inequality test has been improved by choosing the optimal set of measurements \cite{Eberhard_Bell_93}. In future, it is interesting to investigate whether this technique is useful for the proposed QKD scheme.
\end{enumerate}

We thank N.~Beaudry, Q.~Cai, M.~Koashi, H.-K.~Lo, M.~Razavi, and Y.~Tan, for enlightening discussions. This work has been supported by the NSERC Innovation Platform Quantum Works, the NSERC Discovery grant, the FWF (START prize) and the University of Waterloo.

\bibliographystyle{apsrev}

\bibliography{Bibli}

\begin{thebibliography}{33}
\expandafter\ifx\csname natexlab\endcsname\relax\def\natexlab#1{#1}\fi
\expandafter\ifx\csname bibnamefont\endcsname\relax
  \def\bibnamefont#1{#1}\fi
\expandafter\ifx\csname bibfnamefont\endcsname\relax
  \def\bibfnamefont#1{#1}\fi
\expandafter\ifx\csname citenamefont\endcsname\relax
  \def\citenamefont#1{#1}\fi
\expandafter\ifx\csname url\endcsname\relax
  \def\url#1{\texttt{#1}}\fi
\expandafter\ifx\csname urlprefix\endcsname\relax\def\urlprefix{URL }\fi
\providecommand{\bibinfo}[2]{#2}
\providecommand{\eprint}[2][]{\url{#2}}

\bibitem[{\citenamefont{Bell}(1964)}]{Bell_Ineq_64}
\bibinfo{author}{\bibfnamefont{J.~S.} \bibnamefont{Bell}},
  \bibinfo{journal}{Physics} \textbf{\bibinfo{volume}{1}}, \bibinfo{pages}{195}
  (\bibinfo{year}{1964}).

\bibitem[{\citenamefont{Freedman and Clauser}(1972)}]{Freedman.Bell.72}
\bibinfo{author}{\bibfnamefont{S.~J.} \bibnamefont{Freedman}} \bibnamefont{and}
  \bibinfo{author}{\bibfnamefont{J.~F.} \bibnamefont{Clauser}},
  \bibinfo{journal}{Phys. Rev. Lett.} \textbf{\bibinfo{volume}{28}},
  \bibinfo{pages}{938} (\bibinfo{year}{1972}).

\bibitem[{\citenamefont{Aspect et~al.}(1981)\citenamefont{Aspect, Grangier, and
  Roger}}]{Aspect.Bell.81}
\bibinfo{author}{\bibfnamefont{A.}~\bibnamefont{Aspect}},
  \bibinfo{author}{\bibfnamefont{P.}~\bibnamefont{Grangier}}, \bibnamefont{and}
  \bibinfo{author}{\bibfnamefont{G.}~\bibnamefont{Roger}},
  \bibinfo{journal}{Phys. Rev. Lett.} \textbf{\bibinfo{volume}{47}},
  \bibinfo{pages}{460} (\bibinfo{year}{1981}).

\bibitem[{\citenamefont{Tittel et~al.}(1998)\citenamefont{Tittel, Brendel,
  Zbinden, and Gisin}}]{Tittel.Bell.98}
\bibinfo{author}{\bibfnamefont{W.}~\bibnamefont{Tittel}},
  \bibinfo{author}{\bibfnamefont{J.}~\bibnamefont{Brendel}},
  \bibinfo{author}{\bibfnamefont{H.}~\bibnamefont{Zbinden}}, \bibnamefont{and}
  \bibinfo{author}{\bibfnamefont{N.}~\bibnamefont{Gisin}},
  \bibinfo{journal}{Phys. Rev. Lett.} \textbf{\bibinfo{volume}{81}},
  \bibinfo{pages}{3563} (\bibinfo{year}{1998}).

\bibitem[{\citenamefont{Weihs et~al.}(1998)\citenamefont{Weihs, Jennewein,
  Simon, Weinfurter, and Zeilinger}}]{Weihs.Bell.98}
\bibinfo{author}{\bibfnamefont{G.}~\bibnamefont{Weihs}},
  \bibinfo{author}{\bibfnamefont{T.}~\bibnamefont{Jennewein}},
  \bibinfo{author}{\bibfnamefont{C.}~\bibnamefont{Simon}},
  \bibinfo{author}{\bibfnamefont{H.}~\bibnamefont{Weinfurter}},
  \bibnamefont{and}
  \bibinfo{author}{\bibfnamefont{A.}~\bibnamefont{Zeilinger}},
  \bibinfo{journal}{Phys. Rev. Lett.} \textbf{\bibinfo{volume}{81}},
  \bibinfo{pages}{5039} (\bibinfo{year}{1998}).

\bibitem[{\citenamefont{Pan et~al.}(2000)\citenamefont{Pan, Bouwmeester,
  Daniell, Weinfurter, and Zeilinger}}]{Pan.BellGHZ.00}
\bibinfo{author}{\bibfnamefont{J.-W.} \bibnamefont{Pan}},
  \bibinfo{author}{\bibfnamefont{D.}~\bibnamefont{Bouwmeester}},
  \bibinfo{author}{\bibfnamefont{M.}~\bibnamefont{Daniell}},
  \bibinfo{author}{\bibfnamefont{H.}~\bibnamefont{Weinfurter}},
  \bibnamefont{and}
  \bibinfo{author}{\bibfnamefont{A.}~\bibnamefont{Zeilinger}},
  \bibinfo{journal}{Nature} \textbf{\bibinfo{volume}{403}},
  \bibinfo{pages}{515} (\bibinfo{year}{2000}).

\bibitem[{\citenamefont{Rowe et~al.}(2001)\citenamefont{Rowe, Kielpinski,
  Meyer, Sackett, Itano, Monroe, and Wineland}}]{Rowe.Bell.01}
\bibinfo{author}{\bibfnamefont{M.~A.} \bibnamefont{Rowe}},
  \bibinfo{author}{\bibfnamefont{D.}~\bibnamefont{Kielpinski}},
  \bibinfo{author}{\bibfnamefont{V.}~\bibnamefont{Meyer}},
  \bibinfo{author}{\bibfnamefont{C.~A.} \bibnamefont{Sackett}},
  \bibinfo{author}{\bibfnamefont{W.~M.} \bibnamefont{Itano}},
  \bibinfo{author}{\bibfnamefont{C.}~\bibnamefont{Monroe}}, \bibnamefont{and}
  \bibinfo{author}{\bibfnamefont{D.~J.} \bibnamefont{Wineland}},
  \bibinfo{journal}{Nature} \textbf{\bibinfo{volume}{409}},
  \bibinfo{pages}{791} (\bibinfo{year}{2001}).

\bibitem[{\citenamefont{Bohm and Aharonov}(1957)}]{Bohm.Local.57}
\bibinfo{author}{\bibfnamefont{D.}~\bibnamefont{Bohm}} \bibnamefont{and}
  \bibinfo{author}{\bibfnamefont{Y.}~\bibnamefont{Aharonov}},
  \bibinfo{journal}{Phys. Rev.} \textbf{\bibinfo{volume}{108}},
  \bibinfo{pages}{1070} (\bibinfo{year}{1957}).

\bibitem[{\citenamefont{Pearle}(1970)}]{Pearle_Bell_70}
\bibinfo{author}{\bibfnamefont{P.~M.} \bibnamefont{Pearle}},
  \bibinfo{journal}{Phys. Rev. D} \textbf{\bibinfo{volume}{2}},
  \bibinfo{pages}{1418} (\bibinfo{year}{1970}).

\bibitem[{\citenamefont{Curty et~al.}(2004)\citenamefont{Curty, Lewenstein, and
  L\"utkenhaus}}]{CLL_Precondition_04}
\bibinfo{author}{\bibfnamefont{M.}~\bibnamefont{Curty}},
  \bibinfo{author}{\bibfnamefont{M.}~\bibnamefont{Lewenstein}},
  \bibnamefont{and}
  \bibinfo{author}{\bibfnamefont{N.}~\bibnamefont{L\"utkenhaus}},
  \bibinfo{journal}{Phys.~Rev.~Lett.~} \textbf{\bibinfo{volume}{92}},
  \bibinfo{pages}{217903} (\bibinfo{year}{2004}).

\bibitem[{\citenamefont{Mayers}(2001)}]{Mayers_01}
\bibinfo{author}{\bibfnamefont{D.}~\bibnamefont{Mayers}},
  \bibinfo{journal}{Journal of the ACM} \textbf{\bibinfo{volume}{48}},
  \bibinfo{pages}{351–406} (\bibinfo{year}{2001}).

\bibitem[{\citenamefont{Lo and Chau}(1999)}]{LoChauQKD_99}
\bibinfo{author}{\bibfnamefont{H.-K.} \bibnamefont{Lo}} \bibnamefont{and}
  \bibinfo{author}{\bibfnamefont{H.-F.} \bibnamefont{Chau}},
  \bibinfo{journal}{Science} \textbf{\bibinfo{volume}{283}},
  \bibinfo{pages}{2050} (\bibinfo{year}{1999}).

\bibitem[{\citenamefont{Shor and Preskill}(2000)}]{ShorPreskill_00}
\bibinfo{author}{\bibfnamefont{P.~W.} \bibnamefont{Shor}} \bibnamefont{and}
  \bibinfo{author}{\bibfnamefont{J.}~\bibnamefont{Preskill}},
  \bibinfo{journal}{Phys.~Rev.~Lett.~} \textbf{\bibinfo{volume}{85}},
  \bibinfo{pages}{441} (\bibinfo{year}{2000}).

\bibitem[{\citenamefont{Koashi and Preskill}(2003)}]{KoashiPreskill_03}
\bibinfo{author}{\bibfnamefont{M.}~\bibnamefont{Koashi}} \bibnamefont{and}
  \bibinfo{author}{\bibfnamefont{J.}~\bibnamefont{Preskill}},
  \bibinfo{journal}{Phys.~Rev.~Lett.~} \textbf{\bibinfo{volume}{90}},
  \bibinfo{pages}{057902} (\bibinfo{year}{2003}).

\bibitem[{\citenamefont{Gottesman et~al.}(2004)\citenamefont{Gottesman, Lo,
  L\"utkenhaus, and Preskill}}]{GLLP_04}
\bibinfo{author}{\bibfnamefont{D.}~\bibnamefont{Gottesman}},
  \bibinfo{author}{\bibfnamefont{H.-K.} \bibnamefont{Lo}},
  \bibinfo{author}{\bibfnamefont{N.}~\bibnamefont{L\"utkenhaus}},
  \bibnamefont{and} \bibinfo{author}{\bibfnamefont{J.}~\bibnamefont{Preskill}},
  \bibinfo{journal}{Quant. Inf. Comput.} \textbf{\bibinfo{volume}{4}},
  \bibinfo{pages}{325} (\bibinfo{year}{2004}).

\bibitem[{\citenamefont{Makarov et~al.}(2006)\citenamefont{Makarov, Anisimov,
  and Skaar}}]{MAS_Eff_06}
\bibinfo{author}{\bibfnamefont{V.}~\bibnamefont{Makarov}},
  \bibinfo{author}{\bibfnamefont{A.}~\bibnamefont{Anisimov}}, \bibnamefont{and}
  \bibinfo{author}{\bibfnamefont{J.}~\bibnamefont{Skaar}},
  \bibinfo{journal}{Phys. Rev. A} \textbf{\bibinfo{volume}{74}},
  \bibinfo{pages}{022313} (\bibinfo{year}{2006}).

\bibitem[{\citenamefont{Qi et~al.}(2007{\natexlab{a}})\citenamefont{Qi, Fung,
  Lo, and Ma}}]{QFLM_TimeShift_07}
\bibinfo{author}{\bibfnamefont{B.}~\bibnamefont{Qi}},
  \bibinfo{author}{\bibfnamefont{C.-H.~F.} \bibnamefont{Fung}},
  \bibinfo{author}{\bibfnamefont{H.-K.} \bibnamefont{Lo}}, \bibnamefont{and}
  \bibinfo{author}{\bibfnamefont{X.}~\bibnamefont{Ma}},
  \bibinfo{journal}{Quantum Information and Computation}
  \textbf{\bibinfo{volume}{7}}, \bibinfo{pages}{073}
  (\bibinfo{year}{2007}{\natexlab{a}}).

\bibitem[{\citenamefont{Zhao et~al.}(2008)\citenamefont{Zhao, Fung, Qi, Chen,
  and Lo}}]{Timeshift_Exp_08}
\bibinfo{author}{\bibfnamefont{Y.}~\bibnamefont{Zhao}},
  \bibinfo{author}{\bibfnamefont{C.-H.~F.} \bibnamefont{Fung}},
  \bibinfo{author}{\bibfnamefont{B.}~\bibnamefont{Qi}},
  \bibinfo{author}{\bibfnamefont{C.}~\bibnamefont{Chen}}, \bibnamefont{and}
  \bibinfo{author}{\bibfnamefont{H.-K.} \bibnamefont{Lo}},
  \bibinfo{journal}{Phys. Rev. A} \textbf{\bibinfo{volume}{78}},
  \bibinfo{pages}{042333} (\bibinfo{year}{2008}).

\bibitem[{\citenamefont{Nielsen et~al.}(2001)\citenamefont{Nielsen, Schori,
  Sorensen, Salvail, Damgard, and Polzik}}]{Nielsen_Fourstate_01}
\bibinfo{author}{\bibfnamefont{P.~M.} \bibnamefont{Nielsen}},
  \bibinfo{author}{\bibfnamefont{C.}~\bibnamefont{Schori}},
  \bibinfo{author}{\bibfnamefont{J.~L.} \bibnamefont{Sorensen}},
  \bibinfo{author}{\bibfnamefont{L.}~\bibnamefont{Salvail}},
  \bibinfo{author}{\bibfnamefont{I.}~\bibnamefont{Damgard}}, \bibnamefont{and}
  \bibinfo{author}{\bibfnamefont{E.}~\bibnamefont{Polzik}},
  \bibinfo{journal}{Journal of Modern Optics} \textbf{\bibinfo{volume}{48}},
  \bibinfo{pages}{1921} (\bibinfo{year}{2001}).

\bibitem[{\citenamefont{Fung et~al.}(2008)\citenamefont{Fung, Tamaki, Qi, Lo,
  and Ma}}]{Mismatch_security_08}
\bibinfo{author}{\bibfnamefont{C.-H.~F.} \bibnamefont{Fung}},
  \bibinfo{author}{\bibfnamefont{K.}~\bibnamefont{Tamaki}},
  \bibinfo{author}{\bibfnamefont{B.}~\bibnamefont{Qi}},
  \bibinfo{author}{\bibfnamefont{H.-K.} \bibnamefont{Lo}}, \bibnamefont{and}
  \bibinfo{author}{\bibfnamefont{X.}~\bibnamefont{Ma}},
  \bibinfo{journal}{arXiv:0802.3788v1}  (\bibinfo{year}{2008}).

\bibitem[{\citenamefont{Qi et~al.}(2007{\natexlab{b}})\citenamefont{Qi, Fung,
  Zhao, Ma, Tamaki, Chen, and Lo}}]{QFZMTCL_attack_07}
\bibinfo{author}{\bibfnamefont{B.}~\bibnamefont{Qi}},
  \bibinfo{author}{\bibfnamefont{C.-H.~F.} \bibnamefont{Fung}},
  \bibinfo{author}{\bibfnamefont{Y.}~\bibnamefont{Zhao}},
  \bibinfo{author}{\bibfnamefont{X.}~\bibnamefont{Ma}},
  \bibinfo{author}{\bibfnamefont{K.}~\bibnamefont{Tamaki}},
  \bibinfo{author}{\bibfnamefont{C.}~\bibnamefont{Chen}}, \bibnamefont{and}
  \bibinfo{author}{\bibfnamefont{H.-K.} \bibnamefont{Lo}}, in
  \emph{\bibinfo{booktitle}{Quantum Communications and Quantum Imaging V}},
  edited by \bibinfo{editor}{\bibfnamefont{R.~E.} \bibnamefont{Meyers}},
  \bibinfo{editor}{\bibfnamefont{Y.}~\bibnamefont{Shih}}, \bibnamefont{and}
  \bibinfo{editor}{\bibfnamefont{K.~S.} \bibnamefont{Deacon}}
  (\bibinfo{organization}{SPIE}, \bibinfo{year}{2007}{\natexlab{b}}), vol.
  \bibinfo{volume}{6710}, p. \bibinfo{pages}{67100I}.

\bibitem[{\citenamefont{Koashi}(2006)}]{Koashi_NewModel_06}
\bibinfo{author}{\bibfnamefont{M.}~\bibnamefont{Koashi}},
  \bibinfo{journal}{arXiv:quant-ph/0609180}  (\bibinfo{year}{2006}).

\bibitem[{\citenamefont{Tsurumaru and Tamaki}(2008)}]{TT_Thres_08}
\bibinfo{author}{\bibfnamefont{T.}~\bibnamefont{Tsurumaru}} \bibnamefont{and}
  \bibinfo{author}{\bibfnamefont{K.}~\bibnamefont{Tamaki}},
  \bibinfo{journal}{Phys. Rev. A} \textbf{\bibinfo{volume}{78}},
  \bibinfo{pages}{032302} (\bibinfo{year}{2008}).

\bibitem[{\citenamefont{Beaudry et~al.}(2008)\citenamefont{Beaudry, Moroder,
  and L\"utkenhaus}}]{BML_Squash_08}
\bibinfo{author}{\bibfnamefont{N.~J.} \bibnamefont{Beaudry}},
  \bibinfo{author}{\bibfnamefont{T.}~\bibnamefont{Moroder}}, \bibnamefont{and}
  \bibinfo{author}{\bibfnamefont{N.}~\bibnamefont{L\"utkenhaus}},
  \bibinfo{journal}{Phys. Rev. Lett.} \textbf{\bibinfo{volume}{101}},
  \bibinfo{pages}{093601} (\bibinfo{year}{2008}).

\bibitem[{\citenamefont{Ma et~al.}(2007)\citenamefont{Ma, Fung, and
  Lo}}]{EntanglementPDC_07}
\bibinfo{author}{\bibfnamefont{X.}~\bibnamefont{Ma}},
  \bibinfo{author}{\bibfnamefont{C.-H.~F.} \bibnamefont{Fung}},
  \bibnamefont{and} \bibinfo{author}{\bibfnamefont{H.-K.} \bibnamefont{Lo}},
  \bibinfo{journal}{Phys.~Rev.~A} \textbf{\bibinfo{volume}{76}},
  \bibinfo{pages}{012307} (\bibinfo{year}{2007}).

\bibitem[{\citenamefont{L\"utkenhaus}(1999)}]{Lutkenhaus_99DoubleClick}
\bibinfo{author}{\bibfnamefont{N.}~\bibnamefont{L\"utkenhaus}},
  \bibinfo{journal}{Appl.~Phys.~B} \textbf{\bibinfo{volume}{69}},
  \bibinfo{pages}{395} (\bibinfo{year}{1999}).

\bibitem[{\citenamefont{Hwang}(2003)}]{Hwang_03}
\bibinfo{author}{\bibfnamefont{W.-Y.} \bibnamefont{Hwang}},
  \bibinfo{journal}{Phys.~Rev.~Lett.~} \textbf{\bibinfo{volume}{91}},
  \bibinfo{pages}{057901} (\bibinfo{year}{2003}).

\bibitem[{\citenamefont{Lo et~al.}(2005)\citenamefont{Lo, Ma, and
  Chen}}]{Decoy_05}
\bibinfo{author}{\bibfnamefont{H.-K.} \bibnamefont{Lo}},
  \bibinfo{author}{\bibfnamefont{X.}~\bibnamefont{Ma}}, \bibnamefont{and}
  \bibinfo{author}{\bibfnamefont{K.}~\bibnamefont{Chen}},
  \bibinfo{journal}{Phys.~Rev.~Lett.~} \textbf{\bibinfo{volume}{94}},
  \bibinfo{pages}{230504} (\bibinfo{year}{2005}).

\bibitem[{\citenamefont{Duan et~al.}(2001)\citenamefont{Duan, Lukin, Cirac, and
  Zoller}}]{DLCZ_01}
\bibinfo{author}{\bibfnamefont{L.-M.} \bibnamefont{Duan}},
  \bibinfo{author}{\bibfnamefont{M.~D.} \bibnamefont{Lukin}},
  \bibinfo{author}{\bibfnamefont{J.~I.} \bibnamefont{Cirac}}, \bibnamefont{and}
  \bibinfo{author}{\bibfnamefont{P.}~\bibnamefont{Zoller}},
  \bibinfo{journal}{Nature} \textbf{\bibinfo{volume}{414}}, \bibinfo{pages}{413
  } (\bibinfo{year}{2001}).

\bibitem[{\citenamefont{Gottesman and Lo}(2003)}]{TwoWay_03}
\bibinfo{author}{\bibfnamefont{D.}~\bibnamefont{Gottesman}} \bibnamefont{and}
  \bibinfo{author}{\bibfnamefont{H.-K.} \bibnamefont{Lo}},
  \bibinfo{journal}{IEEE Transactions on Information Theory}
  \textbf{\bibinfo{volume}{49}}, \bibinfo{pages}{457} (\bibinfo{year}{2003}).

\bibitem[{\citenamefont{Ma et~al.}(2006)\citenamefont{Ma, Fung, Dupuis, Chen,
  Tamaki, and Lo}}]{TwoWay_06}
\bibinfo{author}{\bibfnamefont{X.}~\bibnamefont{Ma}},
  \bibinfo{author}{\bibfnamefont{C.-H.~F.} \bibnamefont{Fung}},
  \bibinfo{author}{\bibfnamefont{F.}~\bibnamefont{Dupuis}},
  \bibinfo{author}{\bibfnamefont{K.}~\bibnamefont{Chen}},
  \bibinfo{author}{\bibfnamefont{K.}~\bibnamefont{Tamaki}}, \bibnamefont{and}
  \bibinfo{author}{\bibfnamefont{H.-K.} \bibnamefont{Lo}},
  \bibinfo{journal}{Phys.~Rev.~A} \textbf{\bibinfo{volume}{74}},
  \bibinfo{pages}{032330} (\bibinfo{year}{2006}).

\bibitem[{\citenamefont{Mermin}(1986)}]{Mermin_Bell_86}
\bibinfo{author}{\bibfnamefont{N.~D.} \bibnamefont{Mermin}}, in
  \emph{\bibinfo{booktitle}{New Techniques and Ideas in Quantum Measurement
  Theory}}, edited by \bibinfo{editor}{\bibfnamefont{D.~M.}
  \bibnamefont{Greenberger}} (\bibinfo{organization}{New York Academy of
  Science}, \bibinfo{year}{1986}), p. \bibinfo{pages}{422}.

\bibitem[{\citenamefont{Eberhard}(1993)}]{Eberhard_Bell_93}
\bibinfo{author}{\bibfnamefont{P.~H.} \bibnamefont{Eberhard}},
  \bibinfo{journal}{Phys. Rev. A} \textbf{\bibinfo{volume}{47}},
  \bibinfo{pages}{R747} (\bibinfo{year}{1993}).

\end{thebibliography}


\end{document}